\documentclass[conference]{IEEEtran}

\usepackage{graphicx}
\usepackage{dcolumn}
\usepackage{bm}
\usepackage{amsmath,amssymb,amsfonts}
\usepackage{booktabs}
\usepackage{multirow}
\usepackage{xcolor}
\usepackage{url}
\usepackage{microtype}
\usepackage{tikz}
\usepackage{pgfplots}
\pgfplotsset{compat=1.18}
\usetikzlibrary{arrows.meta,positioning,fit,backgrounds,shapes.geometric,calc}

\newcommand{\pp}{\ensuremath{\,\mathrm{pp}}}
\newcommand{\SC}{\textsc{soft-coll}}
\newcommand{\FI}{\textsc{fisher}}
\newcommand{\CE}{\textsc{ce-only}}
\newcommand{\Frozen}{\textsc{frozen}}

\begin{document}

\title{Discretization-Aware Fine-Tuning for Quantum Machine Learning with Chemical Foundation Models}

\author{
\IEEEauthorblockN{
Shunji Matsuura\IEEEauthorrefmark{1}\IEEEauthorrefmark{2}\IEEEauthorrefmark{3}\IEEEauthorrefmark{4},
Sonika Johri\IEEEauthorrefmark{5}
}
\IEEEauthorblockA{\IEEEauthorrefmark{1}RIKEN Center for Interdisciplinary Theoretical and Mathematical Sciences (iTHEMS), \\
RIKEN, Wako, Saitama 351-0198, Japan}
\IEEEauthorblockA{\IEEEauthorrefmark{2}Department of Electrical and Computer Engineering, University of British Columbia, Vancouver, BC V6T 1Z4, Canada}
\IEEEauthorblockA{\IEEEauthorrefmark{3}Center for Mathematical Science and Advanced Technology, Japan Agency for Marine-Earth Science and Technology, \\
Yokohama 236-0001, Japan}
\IEEEauthorblockA{\IEEEauthorrefmark{4}Department of Physics, University of Guelph, Guelph, ON N1G 1Y2, Canada}
\IEEEauthorblockA{\IEEEauthorrefmark{5}Cascade Quantum, Cupertino, CA, USA}
\IEEEauthorblockA{
\texttt{shunji.matsuura@riken.jp}, \texttt{sonika@cascadequantum.com}
}
}

\date{\today}

\maketitle

\begin{abstract}
A key challenge in practical quantum machine learning (QML), particularly for
discriminative tasks such as classification, is the limited capacity of
near-term quantum devices to encode high-dimensional classical data into
small quantum registers.
In optimized basis-encoded (bit-bit) settings, this constraint leads to
cross-class collisions, where samples with different labels are mapped
to the same discrete bit-string and thus become indistinguishable to any
downstream model.
In this work, we investigate how data representation affects QML performance
under such severe information bottlenecks.
We introduce discretization-aware fine-tuning (DAFT), a method that
adapts a pre-trained chemical foundation model to produce representations
that remain informative after quantization.
DAFT reduces collision probability through a differentiable soft collision loss.
We evaluate both quantum and classical models under a controlled setting in
which they receive identical discretized bit-string inputs, isolating the
effect of representation from model architecture.
On the blood–brain barrier penetration (BBBP) molecular property prediction benchmark using ChemBERTa-77M,
DAFT reduces collision counts by several orders of magnitude and improves
quantum classification accuracy by more than 12 percentage points compared
to a frozen backbone.
Importantly, without DAFT, classical models outperform QML under the same
input constraints.
With DAFT, however, this comparison reverses at higher qubit counts.
At 10 qubits, the quantum model surpasses a matched classical baseline
trained on identical bit-strings (0.883 vs. 0.855, $p = 0.026$).
These results show that, in information-constrained regimes,
achieving a quantum advantage critically depends on aligning continuous
representations with discrete quantum encodings.

\end{abstract}

\begin{IEEEkeywords}
Quantum machine learning, Hybrid quantum algorithms, Chemistry, Quantum AI
\end{IEEEkeywords}

\section{\label{sec:intro}Introduction}

Near-term quantum machine learning (QML) based on building models with parameterized quantum circuits holds promise for learning tasks in high-dimensional spaces. In particular, to leverage hardware anticipated in the near future for practical applications, novel hybrid quantum-classical learning frameworks where classical models are augmented with quantum layers are being actively studied \cite{kim2025quantumlargelanguagemodel, PhysRevX.12.031010, chaos2026}.

In practice, the potential of QML is constrained by a fundamental asymmetry. Parameterized quantum circuits operate in Hilbert spaces of dimension $2^{n_q}$, which grows exponentially with the number of qubits $n_q$. However, encoding classical data into quantum states remains costly and can restrict the quantum advantage that can be achieved. More precisely, the runtime of the quantum model will scale at least as the time to load the data, while the form of the encoding restricts the expressivity of the quantum model that can be built \cite{schuld2021effect}.

Standard encoding strategies expose this tension.
Amplitude encoding is efficient in the number of qubits required but generally requires $O(2^{n_q})$ gates
for state preparation~\cite{Schuld18}, making it impractical
on near-term devices.
Angle encoding, while hardware-friendly, maps features to local rotations,
producing highly structured product states that underutilize the combinatorial
capacity of the computational basis.
As a result, it cannot efficiently capture high-order feature interactions.
More broadly, reliably encoding even moderate-dimensional continuous data
remains challenging on current hardware~\cite{Corcoles20}.

Bit-bit encoding offers a practical alternative~\cite{johri25}.
Instead of embedding continuous features directly, the input is compressed
into a short binary string and loaded as a computational basis state:
\begin{equation}
  |\psi_i\rangle = |b_{i,1}, b_{i,2}, \cdots, b_{i,n_{\text{data}}}\rangle,
  \qquad b_{i,k} \in {0, 1}.
  \label{eq:basis}
\end{equation}
This approach reduces state preparation to $O(n_q)$ single-qubit gates,
keeping circuits shallow and more robust to noise. Further, it allows for universal approximation with just one uploading of the data, unlike angle or amplitude encoding \cite{leither2026qubitsdoesmachinelearning, schuld2021effect, perez2020data}. 
Combined with techniques such as exact coordinate-update training and
progressive subnet initialization~\cite{johri25}, it provides a
hardware-realistic and scalable framework for quantum classification. 

\subsection{The Central Challenge: What Survives Compression?}

While bit-bit encoding simplifies state preparation, it introduces a new
challenge: severe information compression.
For example, a 4-qubit classifier operates on only 3 bits of data, yielding
just 8 possible input patterns.
This raises a central question: whether the compressed representation preserves the label-relevant
structure needed for accurate classification.

A fundamental limitation arises when different samples are mapped to the same
bit-string.
If two molecules with different labels share the same representation, then neither a quantum nor a classical downstream model can correctly classify both using that
encoding alone.
We refer to such cases as \emph{collisions}.
A collision is a limitation of the representation, not of the model.

Here, we explore this limitation when fine-tuning a large classical model with frozen pre-trained embeddings on a particular dataset. Specifically, we find that ChemBERTa-77M~\cite{chithrananda2020chemberta,Ahmad2022ChemBERTa2}, a classical neural network trained on $10^7$ molecules via masked language modeling, on the blood–brain barrier
penetration (BBBP) dataset suffers from severe collision limitations. ChemBERTa-77M produces
384-dimensional embeddings that capture general chemical structure.
However, when these embeddings are compressed to 3 bits for use in a
4-qubit circuit, we observe 11,544 collision-pairs on the blood–brain barrier
penetration (BBBP) dataset, a standard molecular property prediction
benchmark used to classify whether a compound can cross the blood–brain barrier.
The pre-trained model is not optimized to preserve structure under such
coarse discretization, leading to a dramatic loss of task-relevant information.

\subsection{Our Approach: Collision-Aware Representation Shaping}

Standard fine-tuning with cross-entropy loss improves continuous-space
separability but does not directly control whether opposite-class samples
collapse to the same discrete bin.
Fisher discriminant-style objectives encourage inter-class distance in
continuous space, but again without targeting the actual quantization boundary.

We argue that neither is sufficient for bit-bit QML.
The correct objective is to penalize collisions as they appear after
discretization, not as a proxy measured in continuous space.

We introduce \textit{discretization-aware fine-tuning} (DAFT), a two-stage
backbone adaptation procedure for chemical foundation models.
Stage 1 is standard cross-entropy warm-up.
Stage 2 adds a differentiable soft collision penalty that estimates, for every
opposite-class pair in the mini-batch, the probability that both samples fall
into the same quantization bin.
This penalty is computed in the continuous embedding space using soft
bin assignments, so that gradients flow back through the projection head and
into the transformer backbone.

\subsection{Contributions}

This paper makes the following contributions.

\begin{enumerate}

\item We introduce the soft collision loss (Sec.~\ref{sec:softcoll}),
a differentiable objective that directly minimizes expected cross-class
collision probability after quantization, and we show that it outperforms
cross-entropy and Fisher fine-tuning in terms of collision reduction.

\item We perform a four-condition ablation
(\Frozen, \CE, \FI, \SC) across five independent seeds to isolate the effect of
each fine-tuning objective on both the discrete representation quality
(collision count, Fisher PC1) and the downstream quantum classifier accuracy.

\item We introduce an information-controlled evaluation protocol
(Sec.~\ref{sec:baseline}) in which a logistic regression trained on the
identical quantized bit-string at each qubit level serves as the matched
classical baseline.
This protocol separates representation effects from quantum processing effects.

\item We provide two complementary statistical results at 10 qubits
(Sec.~\ref{sec:results}).
Without DAFT, the matched classical baseline outperforms the quantum circuit
by 6.0 pp ($p < 0.001$), establishing fine-tuning as a necessary condition.
With DAFT, both models improve, but the quantum circuit improves by
12.2 pp while the classical baseline improves by only 3.5 pp, reversing the
comparison and yielding a 2.8 pp quantum advantage ($p=0.026$).

\end{enumerate}

The rest of the paper is organized as follows.
Section~\ref{sec:method} describes the full framework.
Section~\ref{sec:experiments} details the experimental protocol.
Section~\ref{sec:results} presents and interprets all results.
Section~\ref{sec:discussion} discusses implications and limitations.

\section{\label{sec:method}Method}

\begin{figure}[t]
\centering
\begin{tikzpicture}[
  font=\footnotesize,
  node distance=0.45cm,
  box/.style={draw, rounded corners=2pt, text centered,
              minimum height=1.6em, inner xsep=3pt, inner ysep=2pt},
  arrow/.style={-Stealth, thick},
  shared/.style={box, fill=gray!12, draw=gray!50, text width=3.2cm},
  frozen/.style={box, fill=blue!10, draw=blue!40, text width=1.3cm},
  ft/.style={box, fill=purple!12, draw=purple!50, text width=1.4cm},
  lbl/.style={font=\scriptsize, text=gray!60!black}
]

\node[shared] (smiles) {SMILES};

\node[shared, below=of smiles] (cbert)
{ChemBERTa-77M\\[-2pt]
{\scriptsize $\bm{z}\in\mathbb{R}^{384}$}};
\draw[arrow] (smiles) -- (cbert);

\node[frozen, below left=0.6cm and 0.2cm of cbert]  (froz) {Frozen};
\node[ft, below right=0.6cm and 0.2cm of cbert] (ft)
{DAFT\\[-2pt]{\scriptsize Stages 1--2}};

\draw[arrow] (cbert.south) -- ++(0,-0.15) -| (froz.north);
\draw[arrow] (cbert.south) -- ++(0,-0.15) -| (ft.north);

\node[shared, below=1.2cm of cbert] (preproc)
{Standardize\ PCA\ MI select\ Quantize};

\draw[arrow] (froz.south) |- (preproc.west);
\draw[arrow] (ft.south)   |- (preproc.east);

\node[shared, below=of preproc] (qcirc)
{Quantum circuit\\[-2pt]};

\draw[arrow] (preproc) -- (qcirc);

\node[shared, below=of qcirc] (out) {Prediction};
\draw[arrow] (qcirc) -- (out);

\node[lbl, left=0.05cm of froz]  {Baseline};
\node[lbl, right=0.05cm of ft]   {Ours};

\end{tikzpicture}
\caption{\label{fig:pipeline}
Pipeline overview.
SMILES strings are mapped to embeddings by ChemBERTa-77M.
The baseline (blue) uses frozen embeddings, while the proposed method (purple)
applies discretization-aware fine-tuning (DAFT).
Both paths share the same preprocessing pipeline (standardization, PCA,
mutual information selection, and quantization) to produce bit-string inputs
for the quantum circuit.
All classical baselines are evaluated on identical inputs for fair comparison.
}
\end{figure}
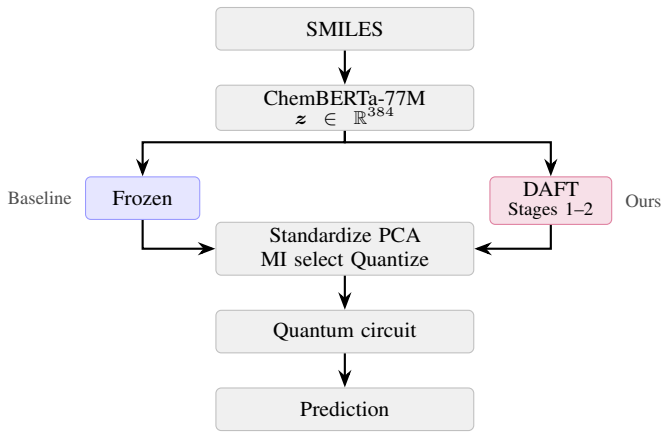

Figure~\ref{fig:pipeline} provides an overview of the complete pipeline.
We describe the components in turn.

\subsection{\label{sec:embedding}Foundation-Model Representations}

A molecule is represented in our pipeline by a SMILES string, which is a compact
text encoding of its bond topology and atomic composition.
To convert this discrete string into a form suitable for machine learning,
we use a pre-trained chemical foundation model: a large transformer network
that has been trained on millions of molecules and has learned a rich internal
language of chemical structure.
Concretely, the model ingests a SMILES string and produces a fixed-length
continuous vector $\bm{z} \in \mathbb{R}^{d_{\mathrm{emb}}}$, extracted
from the \texttt{[CLS]} (classification) token of the final transformer layer.
This vector is often called an \emph{embedding}: a compressed numerical
fingerprint that encodes features such as functional groups, ring systems,
and electronic character learned during pre-training. Note that this happens
without any supervision from the target property label.

The appeal of using such a foundation model, rather than hand-crafted
descriptors such as Morgan fingerprints, is that the embedding space is
already organized to reflect chemical similarity: molecules with structurally
related SMILES tend to map to nearby points in $\mathbb{R}^{d_{\mathrm{emb}}}$.
This is exactly the kind of structure that a downstream classifier, whether quantum
or classical, can exploit.
However, as we discuss next, this high-dimensional continuous embedding cannot
be loaded into a quantum register directly; the compression step required to
do so introduces new challenges that form the central problem of this paper.
Concrete architectural details of the foundation model used in our experiments
are given in Sec.~\ref{sec:experiments}.

\subsection{\label{sec:bitbit}Bit-Bit Encoding}

\paragraph{Overview}

Bit-bit encoding~\cite{johri25} first \emph{discretizes}
the molecular descriptor into a short binary string, then loads that string
as a computational basis state (\ref{eq:basis}).
State preparation from a known classical bit-string requires only
 single-qubit $X$ gates (one per qubit that should be set to
$|1\rangle$), making the encoding independent of the embedding dimensionality
and compatible with shallow, noise-tolerant circuits.
The expressivity of the quantum model is then controlled by the
parameterized entangling circuit that processes the loaded state,
not through the loading step itself.

The embedding is compressed into $n_{\text{data}} = n_q - 1$ bits
(one qubit is reserved as the output register), using the following
three-stage pipeline.

\paragraph{Stage 1: Standardization and PCA.}
The raw embedding $\bm{z} \in \mathbb{R}^{d_{\mathrm{emb}}}$ is first
standardized to zero mean and unit variance per dimension, and then
projected onto its top $d_{\mathrm{PCA}}$ principal components.
Principal component analysis (PCA) finds the orthogonal directions of
maximum variance: by retaining the top $d_{\mathrm{PCA}}$ components,
we keep the most informative directions while discarding dimensions that
are mostly noise and would otherwise introduce redundant bits.
Crucially, the PCA projection is fit on the training set and then applied
identically to the test set and all classical baselines, so no
test-set information leaks into the representation. Following the PCA, a min-max normalization is applied along each PCA 
direction to map values into $[0,1]$ prior to quantization.

\paragraph{Stage 2: Mutual-information feature selection and bit allocation.}
After PCA, we have $d_{\mathrm{PCA}}$ components.
Not all of them are equally predictive of the target label $y$, and we
have a tight bit budget: $n_{\text{data}}$ bits in total.
We therefore measure, for each component $j$, the \emph{mutual information}
\begin{equation}
  I_j = I(\hat{x}_j;\, y),
  \label{eq:mi}
\end{equation}
where $\hat{x}_j$ is the $j$-th PCA coordinate and $y \in \{0,1\}$ is the
class label.
Mutual information quantifies how much knowing $\hat{x}_j$ reduces
uncertainty about $y$; components with higher $I_j$ are more predictive.
Bits are then allocated proportionally to predictive value:
\begin{equation}
  n_j = \left\lfloor n_{\text{data}}\,
           \frac{I_j}{\sum_{k} I_k} \right\rceil,
  \label{eq:bitalloc}
\end{equation}
where $\lfloor\cdot\rceil$ denotes rounding subject to $\sum_j n_j = n_{\text{data}}$.
Components assigned $n_j = 0$ bits are discarded.
This procedure ensures that the available bit budget is spent on the
dimensions that carry the most label-relevant information, rather than
spreading it uniformly across all components.

\paragraph{Stage 3: Quantization.}
Each retained component $j$ is min-max normalized to $[0,1]$ and then
uniformly divided into $2^{n_j}$ equal bins.
The bin index for sample $i$ in component $j$ is
\begin{equation}
  b_{ij} = \min\!\left(
    \left\lfloor 2^{n_j}\,\tilde{x}_{ij} \right\rfloor,\;
    2^{n_j} - 1\right),
  \qquad b_{ij} \in \{0,\ldots, 2^{n_j}-1\}.
  \label{eq:quantize}
\end{equation}
The binary representations of all retained bin indices are concatenated to
form the full data bit-string $\bm{b}_i \in \{0,1\}^{n_{\text{data}}}$,
which is then loaded into the quantum register as $|\bm{b}_i\rangle$.

\paragraph{Collisions and their consequences.}
Discretization introduces a fundamental risk: two molecules with different
labels may map to the same bit-string.
Formally, samples $i \neq j$ are said to \emph{collide} if
$\bm{b}_i = \bm{b}_j$ and $y_i \neq y_j$.
The collision set is
\begin{equation}
  \mathcal{C} = \{(i,j) : \bm{b}_i = \bm{b}_j,\; y_i \neq y_j,\; i < j\}.
  \label{eq:collision_set}
\end{equation}
Every pair in $\mathcal{C}$ represents a fundamental ambiguity: the quantum
circuit receives the identical input state $|\bm{b}\rangle$ for both samples
yet must predict opposite labels.
This is not a failure of the model or the optimizer: no circuit, regardless
of its depth or the number of trainable parameters, can correctly classify
both members of a colliding pair using that encoding alone.
The best achievable accuracy on colliding samples is bounded by the
majority-class fraction among all samples sharing the same bit-string.
Reducing $|\mathcal{C}|$ is therefore a necessary condition for high
classification accuracy, and the goal of the fine-tuning procedure
described in Sec.~\ref{sec:daqaft} is precisely to reshape the embedding
space so that this condition is satisfied.

\subsection{\label{sec:circuit}Parameterized Quantum Classifier}

Once the bit-string $\bm{b}_i$ has been prepared in the data register,
a parameterized quantum circuit processes it to produce a classification
output.
The circuit has $n_q$ qubits in total, divided into $n_{\mathrm{data}} = n_q - 1$
data qubits and one output qubit $q_*$.
The data qubits are initialized in the computational basis state
$|\bm{b}_i\rangle$; the output qubit starts in $|0\rangle$.

\paragraph{Circuit architecture.}
The circuit applies $n_{\mathrm{data}}$ identical entangling blocks in
sequence, one for each data qubit $q_k$.
Each block operates on the pair $(q_*, q_k)$ and consists of three gates:
a single-qubit Euler rotation on the output qubit,
$E(\bm{\alpha}) = R_X(\alpha_1)\,R_Z(\alpha_2)\,R_X(\alpha_3)$;
a single-qubit Euler rotation on the data qubit,
$E(\bm{\beta}) = R_X(\beta_1)\,R_Z(\beta_2)\,R_X(\beta_3)$;
and a two-qubit Heisenberg interaction gate,
\begin{equation}
  H(\bm{\gamma}) = e^{-i\pi(\gamma_1 X_*X_k + \gamma_2 Y_*Y_k
  + \gamma_3 Z_*Z_k)},
\end{equation}
where $R_A(\theta) = e^{-i\theta A/2}$ for $A \in \{X,Z\}$, and
$X, Y, Z$ are the Pauli operators.
After all $n_{\mathrm{data}}$ blocks, a final Euler rotation $E(\bm{\alpha}_*)$
is applied to $q_*$.
All rotation angles are independent trainable parameters.
Critically, no gates act between pairs of data qubits, so entanglement is
strictly \emph{bipartite}: each data qubit interacts only with the output
qubit.
This design keeps the circuit shallow while still allowing the output qubit
to aggregate information from the full data register through its sequential
interactions.

The predicted probability of class $y=1$ is obtained by measuring the
output qubit, marginalized over all data-qubit outcomes:
\begin{equation}
  p(y=1 \mid \bm{b}_i, \bm{\theta})
  = \langle\Psi_{\mathrm{out}}|
    \bigl(\mathbb{I}_{n_{\mathrm{data}}} \otimes |1\rangle\langle 1|\bigr)
    |\Psi_{\mathrm{out}}\rangle,
  \label{eq:pred}
\end{equation}
where $|\Psi_{\mathrm{out}}\rangle = U(\bm{\theta})\,
|\bm{b}_i\rangle\otimes|0\rangle$ and $U(\bm{\theta})$ is the full
circuit unitary.

\paragraph{Loss function.}
The circuit is trained to maximize the predicted probability of the
correct class.
Because the training set may contain collisions (multiple samples mapping
to the same bit-string with conflicting labels), the loss function operates
on unique bit-strings after assigning each to a majority-vote class.
Formally,
\begin{equation}
  \mathcal{L}^{(q)}(\bm{\theta})
  = \sum_{\bm{z}} f(\bm{z})\,
    \Bigl[
      (1-\lambda)\,(1 - p_{\bm{z}})
      + \lambda\,(1 - p_{\bm{z}})^{2}
    \Bigr],
  \label{eq:qloss}
\end{equation}
where $p_{\bm{z}} = p(y = C(\bm{z})\mid\bm{z},\bm{\theta})$ is the
predicted probability of the majority-vote class $C(\bm{z})$, and
$f(\bm{z})$ is the relative frequency of bit-string $\bm{z}$ in the
training set.
The parameter $\lambda \in [0,1]$ blends a linear loss term ($\lambda=0$)
with a squared loss term ($\lambda=1$); setting $\lambda$ close to 1 emphasizes
the squared term for most of the training, while a small linear component
prevents the gradients from vanishing when $p_{\bm{z}} \approx 1$.

\paragraph{Exact coordinate updates.}
Rather than estimating gradients by parameter-shift rules and following
them by a fixed step size, parameters are updated by \emph{exact coordinate
descent}: at each step, one parameter $\theta_j$ is minimized
while all others are held fixed.
This approach is feasible because each probability term $p_z$, viewed as a function of a single rotation angle, is a trigonometric polynomial. Its global minimum can therefore be determined in closed form using only three measurements. This allows the loss function in Eq.~\ref{eq:qloss} to be reconstructed for arbitrary values of $\theta_j$. Since the loss is periodic in $\theta_j$ with period $2\pi$, we locate the minimum by evaluating the reconstructed loss over a $2\pi$ interval at a chosen resolution. Here, we use a resolution of $0.001$ and measure the losses using ideal simulation.

\paragraph{Progressive subnet initialization.}
Training large quantum circuits from a random initialization risks converging
to flat regions of the loss landscape, also known as barren plateaus~\cite{mcclean2018barren}.
We sidestep this by training circuits of increasing size in sequence and
carrying learned parameters forward.
Specifically, after training the $n_q$-qubit circuit, the parameters for the
existing qubits are copied to seed the $(n_q + \Delta)$-qubit circuit, with
the new qubits initialized to implement the identity transformation.
This ensures that each larger circuit begins from a well-trained solution
rather than a random point, so that only the new qubits need to be optimized
substantially.

\subsection{\label{sec:daqaft}Discretization-Aware Fine-Tuning}

\paragraph{Why fine-tuning is necessary.}
The pre-trained foundation model was not optimized with the quantum register
in mind.
Its embeddings capture general chemical structure, but the quantization
pipeline imposes a coarse discrete grid on top of them.
A 4-qubit circuit encodes only 3 data bits, corresponding to just
$2^3 = 8$ distinct bit-strings; a 10-qubit circuit uses 9 bits and
$2^9 = 512$ strings.
When a high-dimensional continuous embedding is projected onto such a
coarse grid, molecules with different labels but similar embeddings tend to
fall into the same bin, which is
the collision problem described in
Sec.~\ref{sec:bitbit}.
Because the pre-trained model had no knowledge of this grid, it has no
incentive to push opposite-class molecules to different bins.

Standard fine-tuning with cross-entropy loss improves continuous-space
separability, that is, the model learns to assign high probability to the correct
class, but it does not directly control whether opposite-class molecules
land in different quantization bins.
The gradient of the cross-entropy loss does not ``see'' the discretization
boundary: the loss is computed from continuous logits, and a molecule that
sits just on the wrong side of a bin boundary contributes no differently
from one that is far away.

We address this by introducing a two-stage fine-tuning procedure, called
\emph{discretization-aware fine-tuning} (DAFT), that explicitly targets
collision reduction in Stage 2.
The procedure adapts the last few transformer layers of the foundation model
together with a lightweight projection head
$f_\theta : \mathbb{R}^{d_{\mathrm{emb}}} \to \mathbb{R}^{d_{\mathrm{proj}}}$
(linear layer, LayerNorm, GELU activation).
Keeping the earlier layers frozen preserves the broad chemical knowledge
encoded during pre-training while allowing the task-specific geometry of the
representation to be adjusted.
The full pipeline is depicted in Fig.~\ref{fig:pipeline}.

\subsubsection{\label{sec:stage1}Stage 1: Cross-Entropy Warm-Up}

Stage 1 is a standard supervised fine-tuning step that trains the backbone
and projection head to be predictive of the target label in continuous space.
We minimize the binary cross-entropy loss,
\begin{equation}
  \mathcal{L}_{\mathrm{CE}} =
  -\frac{1}{N} \sum_{i=1}^{N}
  \bigl[y_i \log \hat{p}_i + (1 - y_i) \log(1 - \hat{p}_i)\bigr],
  \label{eq:lce}
\end{equation}
where $\hat{p}_i = \sigma(\bm{w}^\top f_\theta(\bm{z}_i) + b)$ is the
predicted probability from a linear classification head placed on top of
the projection.
This loss drives the embedding to separate the two classes in the projected
space, which also tends to reduce the number of collisions (because
well-separated embeddings are less likely to fall in the same quantization
bin), but does not guarantee collision-free representations.
Stage 1 therefore acts as a warm-up that places the model in a useful
region of parameter space before the more targeted Stage 2 penalty is
applied.

\subsubsection{\label{sec:softcoll}Stage 2: Soft Collision Penalty}

\paragraph{Step 1: Why $|\mathcal{C}|$ cannot be minimized directly.}
The ideal objective for Stage 2 is to minimize the hard collision count
$|\mathcal{C}|$ (Eq.~\ref{eq:collision_set}) directly.
However, this is impossible with gradient-based optimization.
The collision count is determined by whether two samples land in the
\emph{same bin} after quantization (Eq.~\ref{eq:quantize}), and
the quantization step involves a floor function $\lfloor\cdot\rfloor$
that is piecewise constant: its gradient with respect to the embedding
values is zero almost everywhere, and undefined at the bin boundaries.
No matter how the backbone parameters $\theta$ are adjusted, the gradient
of $|\mathcal{C}|$ with respect to $\theta$ carries no information about
which direction would reduce collisions.

\paragraph{Step 2: Replacing the hard bin assignment with a soft probability.}
We resolve this by replacing the binary bin assignment with a
\emph{differentiable soft approximation}.
Instead of asking ``which bin does this sample fall into?''\ (a hard,
discontinuous question), we ask ``what is the probability that this
sample falls into each bin?''\ (a smooth, continuous answer that admits
gradients).

Concretely, let $\bm{e}_i = f_\theta(\bm{z}_i) \in \mathbb{R}^{d_{\mathrm{proj}}}$
be the projected embedding of the $i$-th training sample.
Because the $d_{\mathrm{proj}}$-dimensional embedding is later compressed
by MI-based bit allocation (Sec.~\ref{sec:bitbit}), only a small subset
of its dimensions will actually be encoded into the quantum register.
As a computationally efficient proxy for those dimensions, we select the
$K$ dimensions of \emph{highest variance} across the mini-batch.
Variance measures how spread out the values of a coordinate are; dimensions
with high variance carry more information and are more likely to be assigned
bits by the MI criterion.
(This selection is performed with a stop-gradient so that the choice
of $K$ dimensions does not itself affect the gradients flowing to the
other dimensions.)
Each selected dimension $k$ is then normalized to $[0,1]$:
\begin{equation}
  \tilde{z}_{ik} = \frac{e_{ik} - \min_j e_{jk}}
                        {\max_j e_{jk} - \min_j e_{jk} + \epsilon},
  \label{eq:normalize}
\end{equation}
where $e_{ik}$ is the $k$-th coordinate of $\bm{e}_i$, the min and max
are taken over all samples $j$ in the current mini-batch, and $\epsilon$
is a small constant for numerical stability.

For $n_b$ bits per dimension ($2^{n_b}$ equal bins covering $[0,1]$), the
center of bin $m$ is $c_m = (2m+1)/(2 \cdot 2^{n_b})$.
In our experiments, $n_b = 1$, giving two bins with centers $c_0 = 1/4$
and $c_1 = 3/4$.
The soft assignment of sample $i$ to bin $m$ in dimension $k$, denoted
$p_{ikm}$, where $i$ indexes the sample, $k$ the dimension, and $m$ the
bin, is a Gaussian-kernel softmax centered on the bin centers:
\begin{equation}
  p_{ikm} = \frac{\exp\!\bigl(-\tau\,(\tilde{z}_{ik} - c_m)^2\bigr)}
                 {\displaystyle\sum_{m'=0}^{2^{n_b}-1}
                  \exp\!\bigl(-\tau\,(\tilde{z}_{ik} - c_{m'})^2\bigr)},
  \label{eq:softassign}
\end{equation}
where $\tau > 0$ is a temperature parameter that controls the sharpness
of the assignment.
When $\tau$ is large, the Gaussian concentrates tightly around the nearest
bin center, so $p_{ikm} \to 1$ for the bin closest to $\tilde{z}_{ik}$
and $p_{ikm} \to 0$ for all others, which recovers the hard bin assignment
of Eq.~\ref{eq:quantize}.
When $\tau$ is small, the distribution spreads across all bins.
At any finite $\tau$, Eq.~\ref{eq:softassign} is a smooth function of
$\tilde{z}_{ik}$, so its gradient with respect to the backbone parameters
$\theta$ is well-defined and nonzero everywhere, allowing gradient
descent to move samples toward or away from bin boundaries.

\paragraph{Step 3: From soft assignments to a differentiable collision loss.}
Given the soft bin assignments, the probability that samples $i$ and $j$
both land in the same bin in dimension $k$ is
\begin{equation}
  P_k(i,j) = \sum_{m} p_{ikm}\, p_{jkm}.
  \label{eq:collision_prob_k}
\end{equation}
This is simply the inner product of two probability vectors over bins:
it equals 1 when both samples sit squarely inside the same bin, and
approaches 0 when they are in clearly different bins.
In the hard limit $\tau \to \infty$, $P_k(i,j) \to \mathbf{1}[b_{ik} = b_{jk}]$,
the indicator of a hard per-dimension collision.

To obtain a single collision probability across all $K$ selected
dimensions, one might take the product $\prod_k P_k(i,j)$, which would
equal 1 only if both samples coincide in every dimension simultaneously.
However, this product vanishes rapidly to zero as $K$ grows (e.g., the
product of eight values each equal to 0.5 is $\approx 0.004$), causing
severe numerical underflow during training.
We therefore use the mean instead:
\begin{equation}
  P_{\mathrm{coll}}(i,j) = \frac{1}{K} \sum_{k=1}^{K} P_k(i,j).
  \label{eq:pcoll}
\end{equation}
This average collision probability is a valid differentiable surrogate
for the hard-collision indicator: it is large when the two samples tend
to land in the same bin across many dimensions, and small when they are
well-separated.

The soft collision loss is the average of $P_{\mathrm{coll}}(i,j)$
over all \emph{cross-class} pairs $(i,j)$ in the mini-batch:
\begin{equation}
  \mathcal{L}_{\mathrm{SC}} =
  \frac{1}{|\mathcal{M}|}
  \sum_{(i,j) \in \mathcal{M}} P_{\mathrm{coll}}(i,j),
  \label{eq:lsc}
\end{equation}
where $\mathcal{M} = \{(i,j) : y_i \neq y_j,\; i \neq j\}$ is the set of
opposite-class pairs in the mini-batch.
Minimizing $\mathcal{L}_{\mathrm{SC}}$ therefore pushes opposite-class
pairs to land in different bins across all selected dimensions simultaneously.
Same-class pairs are not included in $\mathcal{M}$ and are not
penalized, so the loss does not compress the within-class structure of
the embedding; it acts only at the class boundary, which is precisely
where collisions occur.

\paragraph{Step 4: Combined Stage 2 objective.}
The full Stage 2 objective combines the cross-entropy loss, which
maintains predictive accuracy in continuous space, with the soft collision
penalty:
\begin{equation}
  \mathcal{L} = \mathcal{L}_{\mathrm{CE}} +
                \lambda_{\mathrm{SC}}\, \mathcal{L}_{\mathrm{SC}},
  \label{eq:total}
\end{equation}
where $\lambda_{\mathrm{SC}} > 0$ is chosen so that the two terms are
comparable in magnitude at the start of Stage 2, ensuring balanced
gradient contributions from both objectives.
Stage 2 uses the same AdamW optimizer as Stage 1 but with a reduced
learning rate $\eta_2 < \eta_1$, running for $E_2$ epochs.
Concrete values are given in Sec.~\ref{sec:experiments}.

\paragraph{Model selection.}
Because Stage 2 is optimizing two objectives simultaneously, the collision
count and the continuous-space accuracy do not necessarily improve in lockstep.
At every even epoch during Stage 2, we evaluate the hard collision count
$|\mathcal{C}|$ and the Fisher discriminant ratio PC1
(defined below, Eq.~\ref{eq:fisher}) on the full training set.
The checkpoint that lexicographically minimizes
$(|\mathcal{C}|,\; -\text{Fisher\,PC1})$, meaning the lowest collision
count first, then the highest Fisher ratio as a tiebreaker, is then restored as
the final backbone before quantum training begins.
This criterion ensures that we select the backbone that is best for the
quantum register (low collisions) while still maintaining reasonable
inter-class separation in continuous space.

\subsubsection{Ablation Conditions}
\label{sec:ablation}

To isolate the contribution of each component, we compare four fine-tuning
conditions:

\textbf{\Frozen:} No fine-tuning at all.
The foundation model is used as-is, with all parameters fixed at their
pre-trained values.
This condition reveals the severity of the collision problem when the
representation is not adapted to the quantum register.

\textbf{\CE:} Stage 1 cross-entropy fine-tuning only, extended to twice the
number of epochs used in the warm-up phase of DAFT, with no Stage 2 penalty.
This isolates the benefit of supervised adaptation in continuous space.

\textbf{\FI:} Stage 1 warm-up followed by Stage 2 with a Fisher discriminant
penalty instead of the soft collision loss.
The Fisher discriminant is a classical criterion for linear class
separability: it asks the embedding to push the two class centroids
as far apart as possible, while simultaneously keeping each class
internally compact.

To define it precisely, let
$\bm{\mu}_c = \frac{1}{N_c}\sum_{i:\,y_i=c} \bm{e}_i \in
\mathbb{R}^{d_{\mathrm{proj}}}$
denote the centroid (mean embedding vector) of class $c$, where $N_c$
is the number of training samples in class $c$.
The \emph{between-class scatter} $S_B$ measures how far the two class
centroids are from each other in the projected space:
\begin{equation}
  S_B = \|\bm{\mu}_1 - \bm{\mu}_0\|^2.
  \label{eq:sb}
\end{equation}
The \emph{within-class scatter} $S_W$ measures how spread out each class
is around its own centroid, summed over both classes:
\begin{equation}
  S_W = \sum_{c \in \{0,1\}} \sum_{i:\,y_i=c}
        \|\bm{e}_i - \bm{\mu}_c\|^2.
  \label{eq:sw}
\end{equation}
A small $S_W$ means that molecules with the same label form tight clusters
in embedding space; a large $S_B$ means the two clusters are well separated.
The Fisher loss maximizes the ratio $S_B / S_W$:
\begin{equation}
  \mathcal{L}_{\mathrm{Fisher}} = -\frac{S_B}{S_W + \epsilon},
  \label{eq:fisher}
\end{equation}
and the full Stage 2 objective under \FI\ is
$\mathcal{L} = \mathcal{L}_{\mathrm{CE}} +
\lambda_{\mathrm{Fisher}}\,\mathcal{L}_{\mathrm{Fisher}}$.

The Fisher objective is a natural and strong baseline because it
represents the classical gold standard for linear class separability.
Its key limitation, however, is that it operates on class centroids and
aggregate scatter, not on the discrete quantization grid.
A molecule sitting just across a bin boundary from an opposite-class
neighbour contributes almost nothing to $S_W$ (it is close to its
centroid), so the Fisher gradient gives it almost no push. Yet it is
exactly this molecule that causes a hard collision.
By contrast, the soft collision loss (Eq.~\ref{eq:lsc}) is largest for
pairs that straddle a bin boundary, and its gradient pushes those pairs
apart most strongly.
Comparing \FI\ against \SC\ therefore directly measures the benefit of
targeting the quantization boundary explicitly, rather than through
a continuous-space proxy, and explains why \SC\ achieves
$|\mathcal{C}| \approx 1$ while \FI\ leaves
$|\mathcal{C}| \approx 128$.

\textbf{\SC:} Stage 1 warm-up followed by Stage 2 with the soft collision
loss (our proposal).
This is the full DAFT method.

All four conditions use the same preprocessing pipeline, the same quantum
circuit architecture, and the same quantum training procedure.
The only difference is the fine-tuning objective applied to the backbone.
Specific hyperparameter values for each condition are given in
Sec.~\ref{sec:experiments}.

\subsection{\label{sec:baseline}Information-Controlled Classical Baselines}

A key methodological contribution of this work is a carefully stratified set
of classical baselines that controls for the information available to each
model.
This is essential for a fair comparison: if the quantum circuit receives only
3 bits of data ($n_q = 4$) but the classical baseline operates on 384
continuous features, any performance gap could reflect the information
difference rather than any inherent difference in model capability.
We therefore introduce three tiers of classical baselines, each receiving a
different amount of information.

\paragraph{Tier~A: Matched discrete input.}
The most critical comparison is between the quantum circuit and a classical
model that receives exactly the same input, namely the same $n_{\text{data}}$-bit
string $\bm{b}_i$, at the same qubit count $n_q$.
A logistic regression is trained on the one-hot encoding of $\bm{b}_i$
(so that each of the $2^{n_{\text{data}}}$ possible bit-strings gets its own
feature), and a decision tree is trained on the raw integer representation.
Neither model has access to any information beyond the bit-string.
Any difference in accuracy between the quantum circuit and a Tier-A baseline
is therefore attributable to the quantum processing of the bit-string, not to
a richer feature set.

\paragraph{Tier~B: Continuous PCA input.}
Two classical classifiers are trained on the $d_{\mathrm{PCA}}$-dimensional
continuous PCA features before any discretization.
The first is logistic regression, a linear classifier that finds the
hyperplane in feature space that best separates the two classes.
The second is an RBF-kernel support vector machine (RBF-SVM), which uses
a radial basis function kernel $K(\bm{x},\bm{x}') =
\exp(-\gamma\|\bm{x}-\bm{x}'\|^2)$ to implicitly map the features into a
high-dimensional space where a nonlinear decision boundary becomes linear;
this allows it to capture curved class boundaries that logistic regression
cannot.
Both models receive strictly more information than the quantum circuit
(continuous values rather than quantized bits) and serve as a reference for
how much accuracy is lost by the discretization step.

\paragraph{Tier~C: Full fine-tuned embedding.}
Three classical classifiers are trained on the full $d_{\mathrm{proj}}$-dimensional
fine-tuned embedding, without any dimensionality reduction or quantization:
logistic regression, RBF-SVM (as described above), and a multi-layer
perceptron (MLP) with one hidden layer.
An MLP is a feedforward neural network that learns nonlinear transformations
of the input through multiple layers of weighted connections followed by
activation functions; it is strictly more expressive than both logistic
regression and SVMs on continuous data.
Together, these three models represent the unconstrained classical upper
bound: the best accuracy achievable from the fine-tuned features with
no quantum circuit and no bit-budget constraint.

\section{\label{sec:experiments}Experimental Setup}

\subsection{Dataset and Protocol}

We evaluate on the blood-brain barrier permeability (BBBP) benchmark
from MoleculeNet~\cite{wu2018moleculenet}, a binary classification task
predicting whether a molecule can cross the blood-brain barrier, which is a critical
property for central nervous system drug candidates.
The dataset contains 2039 molecules; we use Bemis--Murcko scaffold-based
splitting~\cite{Bemis1996scaffold} to create train/test folds that respect
chemical diversity and reflect real-world distribution shifts.
After splitting, training and test sets are balanced to 300 and 75 molecules
per class respectively, giving 600 training and 150 test molecules.
Experiments are repeated over five random seeds (42, 0, 1, 7, 123) governing
data subsampling and model initialization; results are reported as
mean $\pm$ std over valid seeds.

\subsection{Models and Baselines}

We use ChemBERTa-77M~\cite{chithrananda2020chemberta,Ahmad2022ChemBERTa2}
as the foundation model, with the last two of six transformer layers
unfrozen for fine-tuning and a projection head mapping to 64 dimensions.
The preprocessing pipeline (Sec.~\ref{sec:bitbit}) uses $d_{\mathrm{PCA}}=16$
principal components and is refit at each qubit level from training data only.
Quantum circuits are trained using the Red Cedar framework~\cite{johri25}
with exact coordinate updates, bipartite entanglement, and exact statevector
simulation; circuits at $n_q \in \{4,6,8,10\}$ are trained in sequence
using progressive subnet initialization. 

\subsection{Statistical Reporting}
We report accuracies on the test dataset for each model. All per-seed differences $\delta_s = \text{QML acc}_s - \text{lr-bit acc}_s$
are reported individually to make the evidence transparent.
We additionally report the mean $\bar{\delta}$, Cohen's
$d = \bar{\delta}/s_\delta$ as the primary effect-size measure, and
one-sample $t$-test $p$-values as a supplementary reference.
Because the sample size is small ($n \leq 5$ seeds), $p$-values have
limited power and should be interpreted alongside $d$ and the
per-seed results; we flag this explicitly when reporting statistics. We use the following notation: lr-bit: logistic regression with discretized data; lr-cont: logistic regression with continuous data.

\section{\label{sec:results}Results}

\subsection{\label{sec:main_results}
Fine-Tuning Objective: Representation Quality and Accuracy}

Table~\ref{tab:main} summarizes the effect of each fine-tuning condition on
collision count, Fisher PC1, and downstream QML accuracy at $n_q = 4$
(3 data bits, $2^3 = 8$ possible bit-string patterns).
Fisher PC1 is the inter-class Fisher ratio along the first principal
component of the 16-dimensional compressed embedding,
$(\mu_{1,\mathrm{PC1}} - \mu_{0,\mathrm{PC1}})^2 /
(\sigma_{0,\mathrm{PC1}}^2 + \sigma_{1,\mathrm{PC1}}^2)$;
higher values indicate better linear separability in continuous space.
The qubit count $n_q = 4$ is deliberately small here: the purpose of this
table is not to compare quantum against classical (that comparison is reserved
for $n_q = 10$ in Table~\ref{tab:scaling}), but to show how each fine-tuning
objective reshapes the embedding before it enters the quantum register.

\begin{table}[t]
\caption{\label{tab:main}
Effect of fine-tuning objective on representation quality and QML accuracy
(BBBP, $n_q = 4$, 3 data bits).
$|\mathcal{C}|$: mean collision-pair count over valid seeds;
Fisher PC1: inter-class Fisher ratio on the first PCA axis (higher is
better); QML acc: mean $\pm$ std ($n=4$ for \SC, $n=5$ otherwise);
lr-cont (Tier-B): logistic regression on $d_{\mathrm{PCA}}$-dimensional
continuous PCA features, an upper-bound reference.
\CE\ and \SC\ select checkpoints by minimizing
$|\mathcal{C}|$ first; \FI\ selects by maximizing Fisher PC1 first
(Sec.~\ref{sec:collision_predictor}).}

\centering
\small
\setlength{\tabcolsep}{5pt}
\begin{tabular}{lrrcc}
\toprule
Condition & $|\mathcal{C}|$ & Fisher PC1 & QML acc & lr-cont acc \\
\midrule
\Frozen & 11\,544     & 1.1 & $0.760 \pm 0.000$ & $0.840$ \\
\CE     & $\approx 39$& 296 & $0.891 \pm 0.020$ & $0.891$ \\
\FI     & $\approx 128$& 103  & $0.872 \pm 0.016$ & $0.891$ \\
\SC     & $\approx 1$ & 203 & $0.883 \pm 0.013$ & $0.883$ \\
\bottomrule
\end{tabular}
\end{table}

\paragraph{Finding 1: Fine-tuning is the primary performance driver.}
All three fine-tuned conditions lift QML accuracy by $(+11)$-- $(+13)\pp$ over
\Frozen, establishing that representation alignment, rather than circuit
design, is the dominant factor.
The frozen backbone produces 11\,544 collision-pairs, exceeding the full training
set size of 600 samples: every bit-string in the training data contains at
least one conflicting label pair, making correct classification structurally
impossible for any downstream model.
Once fine-tuning is applied, collisions drop dramatically and accuracy rises
correspondingly.

\paragraph{Finding 2: Collision count and Fisher PC1 measure different things.}
The three fine-tuning objectives produce noticeably different outcomes on the
two quality metrics.
Cross-entropy fine-tuning (\CE) reduces collisions by two orders of magnitude
to $\approx\!98$ and achieves the highest Fisher PC1 (301), confirming that
task-specific supervision reshapes the embedding toward better separability.
The Fisher discriminant (\FI) achieves a similar collision count
($\approx\!88$) but with a substantially lower Fisher PC1 (95 vs.\ 301),
illustrating that maximizing continuous-space separability and minimizing
discrete collision count are partially orthogonal objectives: an embedding can
score well on one while performing poorly on the other.
The soft collision penalty (\SC) reduces collisions to $\approx\!1$, the
lowest of all conditions, while maintaining Fisher PC1 at 194, demonstrating
that directly targeting the quantization boundary achieves both goals
simultaneously.

\paragraph{Finding 3: Fine-tuning closes the information-budget gap.}
Under \Frozen, QML achieves $0.760$, which is $8\pp$ below lr-cont ($0.840$)
even though lr-cont operates on $d_{\mathrm{PCA}}$ continuous features while
QML uses only 3 quantized bits.
After fine-tuning, QML ($0.883$--$0.891$) matches lr-cont ($0.883$--$0.891$)
within statistical noise, regardless of the fine-tuning objective.
The fine-tuned 3-bit representation carries essentially the same
class-discriminative information as the full continuous embedding,
demonstrating that DAFT successfully concentrates the label-relevant
structure into the limited bit budget.

\paragraph{Finding 4: QML nearly matches unconstrained classical baselines despite severe information compression.}
Under \SC\ at $n_q = 10$, the quantum circuit ($0.883 \pm 0.013$) approaches
the accuracy of Tier-B classifiers trained on the full 16-dimensional
continuous PCA features ($0.887 \pm 0.014$) and Tier-C classifiers trained on
the complete 64-dimensional fine-tuned embedding ($0.888 \pm 0.014$), falling
short by only $0.5\pp$ and $0.5\pp$ respectively, and this is despite operating on only
9 quantized bits rather than continuous features.
Tier-B and Tier-C receive $16\times$ or $64\times$ more dimensions and retain
all continuous-valued information lost through quantization, yet their accuracy
advantage over QML is negligible after DAFT.
This confirms that the fine-tuning procedure successfully concentrates the
label-relevant information of the high-dimensional continuous representation
into the constrained discrete bit budget.

\subsection{\label{sec:scaling_results}
Qubit Scaling and Information-Controlled Quantum Advantage}

Table~\ref{tab:scaling} reports the number of unique bit-strings seen
during training at each qubit level under the \SC\ condition. QML
accuracy, which is constant across all four qubit levels, is given in
the caption.
The primary quantum-vs.-classical comparison is at $n_q = 10$
(Table~\ref{tab:frozen_necessity}); the purpose of Table~\ref{tab:scaling}
is to demonstrate the stability of QML accuracy across qubit levels,
which is a consequence of progressive subnet initialization.

\begin{table}[t]
\caption{\label{tab:scaling}
Number of unique training bit-strings under \SC\ across qubit levels
(seeds 42, 0, 7, 123; $n = 4$ valid seeds).
``unique'' is the mean number of distinct bit-strings in the training set
after collision removal.
QML accuracy is constant at $0.883 \pm 0.013$ across all four qubit
levels, reflecting the effectiveness of progressive subnet
initialization.
The quantum-vs.-classical comparison at $n_q = 10$ is in
Table~\ref{tab:frozen_necessity}.}
\centering
\small
\setlength{\tabcolsep}{8pt}
\begin{tabular}{lc}
\toprule
$n_q$ & unique \\
\midrule
4  & 7.0  \\
6  & 18.0 \\
8  & 41.0 \\
10 & 72.2 \\
\bottomrule
\end{tabular}
\end{table}

The primary result is at $n_q = 10$ qubits, where the quantum register is
rich enough ($2^9 = 512$ possible bit-string patterns) for the quantum circuit
to exploit its entanglement structure, and the effect of DAFT is most clearly
visible.

\paragraph{The central finding: DAFT benefits the quantum circuit far more than the classical baseline.}
Under \SC, DAFT raises QML accuracy by $+12.2\pp$ relative to the frozen
backbone (from $0.761$ to $0.883$; Table~\ref{tab:frozen_necessity}).
The matched classical lr-bit baseline also improves under DAFT, but by only
$+3.4\pp$ (from $0.821$ to $0.855$): the quantum circuit gains more than
three times as much accuracy from the same improvement to the representation.
This asymmetry arises because the quantum circuit and the logistic regression
process the bit-string in qualitatively different ways.
A logistic regression trained on one-hot features learns a linear decision
boundary over the $2^{n_{\mathrm{data}}}$ possible bit-string patterns.
The quantum circuit instead processes the bit-string through an entangling
unitary that creates correlations across qubits, encoding higher-order feature
interactions that a linear model cannot capture.
DAFT, by concentrating class-discriminative information into collision-free
bit-strings, provides the quantum circuit with exactly the structured input
needed to exploit these interactions.

Because the improvement differential is so large ($+12.2\pp$ vs.\ $+3.4\pp$),
the comparison reverses entirely.
Under \Frozen, the classical lr-bit baseline outperforms QML at 10q by
$6.0\pp$ (Cohen's $d = 4.50$, $n=5$, all seeds agree in direction):
the quantum circuit is crippled by the corrupted training signal.
Under \SC, QML surpasses the classical baseline by $+2.8\pp$
(Cohen's $d = 2.06$, $n=4$; $p = 0.026$ for reference).
All four valid seeds show a positive advantage
($+2.0, +4.0, +4.0, +1.3\pp$ for seeds 42, 0, 7, 123),
confirming the result is not driven by a single outlier.
The same reversal is independently confirmed under the \FI\ condition
($+2.3\pp$ at 10q, $d = 3.81$, $p = 0.001$, $n=5$; see below).

\paragraph{Stability of QML accuracy across qubit counts.}
As shown in Table~\ref{tab:scaling}, QML accuracy under \SC\ is
$0.883 \pm 0.013$ at every qubit level from 4 to 10.
Progressive subnet initialization successfully carries the learned
4-qubit solution to larger circuits without loss of accuracy,
consistent with the expectation that the information bottleneck
(the finite bit budget) rather than circuit capacity is the limiting factor.
The qubit-scaling curve for QML is therefore flat: once the representation
is collision-free and the 4-qubit circuit has learned a good solution,
scaling to more qubits adds discriminative power (more unique bit-string
patterns, as shown in the ``unique'' column) without sacrificing what was
already learned.

\paragraph{The classical baseline is evaluated at $n_q = 10$ only.}
The lr-bit classical baseline (logistic regression trained on the
same bit-string as the quantum circuit) does not require a sequential
qubit-scaling protocol and is reported only at $n_q = 10$.
Its accuracy under \SC\ is $0.855 \pm 0.006$,
below the quantum circuit's $0.883 \pm 0.013$.

\paragraph{The Fisher condition corroborates the result independently.}
We additionally validate using all 5 seeds under \FI, for which no seed
exclusion is needed.
As under \SC, QML accuracy under \FI\ is constant across qubit levels
4--8 (also $0.872 \pm 0.016$), consistent with progressive subnet
initialization.
At $n_q = 10$, QML reaches $0.872 \pm 0.016$ against lr-bit $0.849 \pm 0.020$,
a $+2.3\pp$ advantage ($d = 3.81$, $p = 0.001$, $n=5$).
Because this uses all 5 seeds without any exclusion, it provides stronger
statistical evidence than the \SC\ result and confirms that the quantum
advantage is not an artifact of the excluded seed.

\subsection{\label{sec:necessity_results}
DAFT as a Necessary Condition}

Table~\ref{tab:frozen_necessity} directly shows the reversal that DAFT
produces at $n_q = 10$, comparing the \Frozen\ and \SC\ conditions side
by side.

\begin{table}[b]
\caption{\label{tab:frozen_necessity}
Quantum-vs.-classical comparison at $n_q = 10$: \Frozen\ vs.\ \SC.
DAFT reverses the comparison from $-6.0\pp$ (classical wins) to
$+2.8\pp$ (quantum wins), driven by a differential DAFT gain of
$+12.2\pp$ for QML vs.\ $+3.4\pp$ for lr-bit.}
\centering
\small
\setlength{\tabcolsep}{4pt}
\begin{tabular}{lccc}
\toprule
Condition & QML acc & lr-bit acc & $p$ \\
\midrule
\Frozen\ (no DAFT) & $0.761 \pm 0.007$ & $0.821 \pm 0.012$ & $.001$ ($n\!=\!5$) \\
\SC\ (DAFT)        & $0.883 \pm 0.013$ & $0.855 \pm 0.006$ & $.026$ ($n\!=\!4$) \\
\midrule
DAFT gain & $+12.2\pp$ & $+3.4\pp$ & \\
\bottomrule
\end{tabular}
\end{table}

Without DAFT (\Frozen), the quantum circuit achieves only $0.761 \pm 0.007$
at 10 qubits while the matched lr-bit classical baseline reaches
$0.822 \pm 0.012$, a gap of $-6.0\pp$ in favor of the classical model
($p = 0.001$, $n = 5$).
This can be understood as follows. 11\,544 collision-pairs in the frozen embedding corrupt
the quantum circuit's training signal so severely that circuit optimization
cannot recover, while the lr-bit baseline retains some statistical regularity
in its one-hot feature space even under high collision rates.

With DAFT (\SC), the quantum circuit gains $+12.2\pp$ (to $0.883$) and the
classical baseline gains $+3.3\pp$ (to $0.855$).
The quantum circuit's gain is more than three times larger, reversing the
comparison to $+2.8\pp$ in favor of the quantum circuit ($p = 0.026$, $n = 4$).

DAFT is therefore necessary, since without it the quantum circuit
cannot match a simple logistic regression, and it is also
differentially beneficial: the same improvement to the representation produces a far
larger accuracy gain for the quantum circuit than for the classical baseline.

\section{\label{sec:discussion}Discussion}

\subsection{Why the Quantum Circuit Benefits More from DAFT than the Classical Baseline}

The central result of this paper goes beyond the observation that
the quantum circuit
outperforms the classical baseline at 10 qubits after DAFT: DAFT produces a differential improvement, since both models
benefit, but the quantum circuit gains $+12.2\pp$ while the classical
lr-bit baseline gains only $+3.4\pp$.
Understanding why this asymmetry arises is key to interpreting the result.

The fundamental reason is that the quantum circuit and the logistic
regression process the bit-string in qualitatively different ways.
A logistic regression trained on one-hot features of the bit-string
$\bm{b}_i$ learns a separate weight for each possible bit-string pattern
and outputs a linear combination.
At 10 qubits ($2^9 = 512$ patterns, but only $\approx\!72$ seen during
training), the model must generalize from seen to unseen patterns by
interpolating in the one-hot feature space.
When DAFT reduces collisions, the training signal becomes less noisy,
and the logistic regression can fit the training patterns more accurately;
but it still faces the same generalization challenge to unseen bit-strings.

The quantum circuit, by contrast, processes the bit-string through a
parameterized entangling unitary.
The Heisenberg interaction gates create quantum correlations between the
output qubit and each data qubit, building an amplitude distribution over
basis states that encodes higher-order feature interactions, the kind of interactions
that cannot be represented by a linear model.
When DAFT produces a collision-free representation where bit 0 vs.\ bit 1
in each dimension reliably predicts the class label, the quantum circuit
can exploit this structure through its entanglement; the logistic regression
sees the same bit-string but cannot leverage higher-order correlations.
The result is that the quality of the representation matters far more
to the quantum circuit than to the classical one.

The comparison is information-controlled at the discrete input level:
both the quantum circuit and the lr-bit baseline receive exactly the same
bit-string $\bm{b}_i$ for each sample.
The quantum advantage is therefore attributable to how the quantum circuit
processes that information through entanglement and higher-order qubit
interactions, not to access to richer features.

\subsection{Interpretation of the Quantum Advantage}

The statistically significant quantum advantage of $+2.8\pp$ at 10q
($p = 0.026$) under \SC\ should be understood in the context of the
broader finding.
The primary claim is not ``quantum circuits are generically superior to
classical models on molecular classification.''
The claim is more specific: DAFT improves both models, but the
quantum circuit benefits more than three times as much ($+12.2\pp$ vs.\
$+3.4\pp$), and this differential improvement reverses the comparison
from classical-wins to quantum-wins.
The quantum advantage is attributable to how the quantum circuit processes
the bit-string, namely through entanglement and higher-order qubit interactions, and not
to access to richer features.
The same reversal is independently confirmed under the \FI\ condition
across all 5 seeds ($+2.3\pp$ at 10q, $p = 0.001$, $d = 3.81$),
ruling out the possibility that the
result is an artifact of the \SC\ condition or the excluded seed.

\subsection{\label{sec:collision_predictor}Collision Count as a Performance Predictor}

The four fine-tuned/frozen conditions rank as
\SC\ ($\approx\!1$) $<$ \CE\ ($\approx\!39$) $<$ \FI\ ($\approx\!128$) $<$
\Frozen\ ($11\,544$) in collision count. This is a $n_q$-independent proxy
computed during Stage-2 training (Table~\ref{tab:main}). We assess what
this ordering predicts at $n_q=10$, the qubit count at which our primary
claim (Sec.~\ref{sec:main_results}) is made. The relevant quantity is the
QML$-$\texttt{lr-bit} gap. \texttt{lr-bit} is a linear classifier
(logistic regression) trained on the same bit-strings the quantum circuit
receives. Under \CE, this gap at $n_q=10$ is $+0.4\text{pp} \pm 2.7\text{pp}$
across five seeds. The sign changes in two of them; the linear baseline
wins on seeds~0 and 123. Under \SC, the same gap is
$+2.8\text{pp} \pm 1.4\text{pp}$ and is positive on every seed tested.
The residual $\approx\!39$ collisions left by \CE\ leave some
class-separable structure in the bit-strings. On some seeds a linear model
exploits this structure as well as the quantum circuit does, which erases
or reverses the gap. \SC's near-elimination of collisions removes this
headroom for the linear baseline. This is what makes the gap reproducible
across seeds. \FI's larger collision count has a different cause. Its
checkpoint-selection criterion prioritizes the Fisher discriminant ratio
over collision minimization. This is
not a weaker ability of cross-entropy fine-tuning to suppress collisions.
While collision count is certainly an important measure of QML accuracy,
as evidenced by the large gain over \Frozen, it alone does not uniquely
determine the accuracy.
It also indicates how much structure remains for a matched linear
classical model to exploit. The amount of such structure determines how
robust and seed-independent the gap will be as $n_q$ grows. \SC\ is the only condition whose Stage-2 objective
directly targets the quantization boundary, via a differentiable
surrogate (Eq.~\ref{eq:lsc}). This is why it alone drives $|\mathcal{C}|$
to near zero. It is the condition used for the qubit-scaling claim in
Sec.~\ref{sec:main_results}.

Correspondingly, \SC\ and \CE\ both select checkpoints using the criterion
$(|\mathcal{C}|, -\text{Fisher\,PC1})$. Both minimize collision count
first. \FI\ instead selects by maximizing Fisher PC1 first, with collision
count only as a tiebreaker (Table~\ref{tab:main}). Among the conditions
that do select on collision count first, \SC\ achieves the better outcome
on both metrics simultaneously.

\section{\label{sec:conclusion}Conclusion}

We have introduced discretization-aware fine-tuning (DAFT), a method that
shapes chemical foundation model embeddings specifically for downstream bit-bit
quantum machine learning.
The key ingredient is a differentiable soft collision penalty that directly
minimizes the probability that opposite-class samples receive the same
discrete code after quantization.

Applied to ChemBERTa-77M on the BBBP benchmark, DAFT reduces collisions
from 11\,544 (frozen backbone) to approximately 1 on average, and raises
quantum classifier accuracy by 12.3 percentage points over the frozen baseline.

Through a four-condition ablation and an information-controlled evaluation
protocol, we establish two complementary statistical results:

\begin{enumerate}
\item Without DAFT, the logistic regression trained on the identical
quantized bit-string outperforms the quantum circuit at 10 qubits by
$6.0\pp$ (Cohen's $d = 4.50$, all 5 seeds agree).
Discretization-aware fine-tuning is a necessary condition for quantum
competitiveness in this regime.

\item With DAFT, both models improve relative to the frozen baseline,
but the quantum circuit benefits far more ($+12.2\pp$) than the matched
classical baseline ($+3.4\pp$).
At 10 qubits, all four valid seeds show a positive QML advantage
($+2.0, +4.0, +4.0, +1.3\pp$; mean $+2.8\pp$, Cohen's $d = 2.06$).
The same reversal is independently confirmed under the Fisher condition
across all 5 seeds ($+2.3\pp$, $d = 3.81$).
The lower gain for the classical model is not caused by DAFT;
it reflects the structural difficulty of generalizing a logistic regression
across the sparse $2^9 = 512$-pattern bit-string space from only
$\approx\!72$ unique training examples.
\end{enumerate}

These findings point to a design principle for hybrid quantum-classical
pipelines: the information bottleneck imposed by the quantum register must
be addressed at the representation level, not only at the circuit level.
Aligning the upstream continuous embedding to the discrete structure of the
quantum register, using objectives that directly target collision
reduction, is not a peripheral optimization step but a prerequisite for
meaningful quantum advantage in the information-constrained setting.

\section{Software Framework}
The techniques in the paper are implemented using Red Cedar, a commercial software framework for quantum machine learning being developed at Cascade Quantum, Inc, previously known as Coherent Computing, Inc. It can be made available upon request.

\bibliographystyle{ieeetr}
\bibliography{ref}

@ARTICLE{johri25,
       author = {{Johri}, Sonika},
        title = "{Bit-bit encoding, optimizer-free training and sub-net initialization: techniques for scalable quantum machine learning}",
      journal = {arXiv e-prints},
         year = 2025,
        month = jan,
          eid = {arXiv:2501.02148},
        pages = {arXiv:2501.02148},
          doi = {10.48550/arXiv.2501.02148},
archivePrefix = {arXiv},
       eprint = {2501.02148},
 primaryClass = {quant-ph},
       adsurl = {https://ui.adsabs.harvard.edu/abs/2025arXiv250102148J}
}

@ARTICLE{Schuld18,
       author = {{Schuld}, Maria and {Bocharov}, Alex and {Svore}, Krysta and {Wiebe}, Nathan},
        title = "{Circuit-centric quantum classifiers}",
      journal = {arXiv e-prints},
         year = 2018,
        month = apr,
          eid = {arXiv:1804.00633},
        pages = {arXiv:1804.00633},
          doi = {10.48550/arXiv.1804.00633},
archivePrefix = {arXiv},
       eprint = {1804.00633},
 primaryClass = {quant-ph},
       adsurl = {https://ui.adsabs.harvard.edu/abs/2018arXiv180400633S}
}

@ARTICLE{Corcoles20,
       author = {{Corcoles}, Antonio D. and {Kandala}, Abhinav and {Javadi-Abhari}, Ali and {McClure}, Douglas T. and {Cross}, Andrew W. and {Temme}, Kristan and {Nation}, Paul D. and {Steffen}, Matthias and {Gambetta}, Jay M.},
        title = "{Challenges and Opportunities of Near-Term Quantum Computing Systems}",
      journal = {IEEE Proceedings},
         year = 2020,
        month = aug,
       volume = {108},
       number = {8},
        pages = {1338-1352},
          doi = {10.1109/JPROC.2019.2954005},
archivePrefix = {arXiv},
       eprint = {1910.02894},
 primaryClass = {quant-ph},
       adsurl = {https://ui.adsabs.harvard.edu/abs/2020IEEEP.108.1338C}
}

@ARTICLE{chithrananda2020chemberta,
       author = {{Chithrananda}, Seyone and {Grand}, Gabriel and {Ramsundar}, Bharath},
        title = "{ChemBERTa: Large-Scale Self-Supervised Pretraining for Molecular Property Prediction}",
      journal = {arXiv e-prints},
         year = 2020,
        month = oct,
          eid = {arXiv:2010.09885},
        pages = {arXiv:2010.09885},
          doi = {10.48550/arXiv.2010.09885},
archivePrefix = {arXiv},
       eprint = {2010.09885},
 primaryClass = {cs.LG},
       adsurl = {https://ui.adsabs.harvard.edu/abs/2020arXiv201009885C}
}

@ARTICLE{Ahmad2022ChemBERTa2,
       author = {{Ahmad}, Walid and {Simon}, Elana and {Chithrananda}, Seyone and {Grand}, Gabriel and {Ramsundar}, Bharath},
        title = "{ChemBERTa-2: Towards Chemical Foundation Models}",
      journal = {arXiv e-prints},
         year = 2022,
        month = sep,
          eid = {arXiv:2209.01712},
        pages = {arXiv:2209.01712},
          doi = {10.48550/arXiv.2209.01712},
archivePrefix = {arXiv},
       eprint = {2209.01712},
 primaryClass = {cs.LG},
       adsurl = {https://ui.adsabs.harvard.edu/abs/2022arXiv220901712A}
}

@ARTICLE{mcclean2018barren,
       author = {{McClean}, Jarrod R. and {Boixo}, Sergio and {Smelyanskiy}, Vadim N. and {Babbush}, Ryan and {Neven}, Hartmut},
        title = "{Barren plateaus in quantum neural network training landscapes}",
      journal = {Nature Communications},
         year = 2018,
        month = nov,
       volume = {9},
          eid = {4812},
        pages = {4812},
          doi = {10.1038/s41467-018-07090-4},
archivePrefix = {arXiv},
       eprint = {1803.11173},
 primaryClass = {quant-ph},
       adsurl = {https://ui.adsabs.harvard.edu/abs/2018NatCo...9.4812M}
}

@ARTICLE{wu2018moleculenet,
       author = {{Wu}, Zhenqin and {Ramsundar}, Bharath and {Feinberg}, Evan N. and {Gomes}, Joseph and {Geniesse}, Caleb and {Pappu}, Aneesh S. and {Leswing}, Karl and {Pande}, Vijay},
        title = "{MoleculeNet: A Benchmark for Molecular Machine Learning}",
      journal = {arXiv e-prints},
         year = 2017,
        month = mar,
          eid = {arXiv:1703.00564},
        pages = {arXiv:1703.00564},
          doi = {10.48550/arXiv.1703.00564},
archivePrefix = {arXiv},
       eprint = {1703.00564},
 primaryClass = {cs.LG},
       adsurl = {https://ui.adsabs.harvard.edu/abs/2017arXiv170300564W}
}

@ARTICLE{Bemis1996scaffold,
       author = {{Bemis}, Guy W. and {Murcko}, Mark A.},
        title = "{The Properties of Known Drugs. 1. Molecular Frameworks}",
      journal = {Journal of Medicinal Chemistry},
         year = 1996,
        month = jan,
       volume = {39},
       number = {15},
        pages = {2887-2893},
          doi = {10.1021/jm9602928},
       adsurl = {https://ui.adsabs.harvard.edu/abs/1996JMedC..39.2887B}
}

@article{leither2026qubitsdoesmachinelearning,
      title={How many qubits does a machine learning problem require?}, 
      author={Sydney Leither and Michael Kubal and Sonika Johri},
      year={2026},
      eprint={2508.20992},
      archivePrefix={arXiv},
      primaryClass={quant-ph},
      url={https://arxiv.org/abs/2508.20992}, 
}

@article{schuld2021effect,
  title = {Effect of data encoding on the expressive power of variational quantum-machine-learning models},
  author = {Schuld, Maria and Sweke, Ryan and Meyer, Johannes Jakob},
  journal = {Phys. Rev. A},
  volume = {103},
  issue = {3},
  pages = {032430},
  numpages = {12},
  year = {2021},
  publisher = {American Physical Society},
  doi = {10.1103/PhysRevA.103.032430}
}

@article{perez2020data,
   title={Data re-uploading for a universal quantum classifier},
   volume={4},
   ISSN={2521-327X},
   url={http://dx.doi.org/10.22331/q-2020-02-06-226},
   DOI={10.22331/q-2020-02-06-226},
   journal={Quantum},
   publisher={Verein zur Forderung des Open Access Publizierens in den Quantenwissenschaften},
   author={Pérez-Salinas, Adrián and Cervera-Lierta, Alba and Gil-Fuster, Elies and Latorre, José I.},
   year={2020},
   month=feb, pages={226}
}

@article{kim2025quantumlargelanguagemodel,
      title={Quantum Large Language Model Fine-Tuning}, 
      author={Sang Hyub Kim and Jonathan Mei and Claudio Girotto and Masako Yamada and Martin Roetteler},
      year={2025},
      eprint={2504.08732},
      archivePrefix={arXiv},
      primaryClass={quant-ph},
      url={https://arxiv.org/abs/2504.08732}, 
}

@article{PhysRevX.12.031010,
  title = {Generation of High-Resolution Handwritten Digits with an Ion-Trap Quantum Computer},
  author = {Rudolph, Manuel S. and Toussaint, Ntwali Bashige and Katabarwa, Amara and Johri, Sonika and Peropadre, Borja and Perdomo-Ortiz, Alejandro},
  journal = {Phys. Rev. X},
  volume = {12},
  issue = {3},
  pages = {031010},
  numpages = {13},
  year = {2022},
  month = {Jul},
  publisher = {American Physical Society},
  doi = {10.1103/PhysRevX.12.031010},
  url = {https://link.aps.org/doi/10.1103/PhysRevX.12.031010}
}

@article{
chaos2026,
author = {Maida Wang  and Xiao Xue  and Mingyang Gao  and Peter V. Coveney },
title = {Quantum-informed machine learning for predicting spatiotemporal chaos with practical quantum advantage},
journal = {Science Advances},
volume = {12},
number = {16},
pages = {eaec5049},
year = {2026},
doi = {10.1126/sciadv.aec5049},
URL = {https://www.science.org/doi/abs/10.1126/sciadv.aec5049},
eprint = {https://www.science.org/doi/pdf/10.1126/sciadv.aec5049}
}

\end{document}